\documentclass[12pt,a4paper]{amsart}

\usepackage{amsmath,amssymb,amsthm,bm}
\usepackage{latexsym}
\usepackage{amsfonts}
\usepackage{bbm,dsfont}
\usepackage{color}

\numberwithin{equation}{section}

\theoremstyle{definition}

\newtheorem{theorem}{Theorem}

\newcommand{\N}{\mathbb N}
\newcommand{\R}{\mathbb R}

\newcommand{\hi}{\mathcal{H}} 
\newcommand{\id}{I}

\newcommand{\lh}{\mathcal{L(H)}} 

\newcommand{\tr}[1]{\mathrm{tr}\left[#1\right]} 

\newcommand{\Povms}{\mathcal{P}(\Omega,\Sigma,\hi)} 
\newcommand{\Ao}{\mathsf{A}} 
\newcommand{\Bo}{\mathsf{B}} 
\newcommand{\Co}{\mathsf{C}} 
\newcommand{\Eo}{\mathsf{E}} 

\def\<{\langle}
\def\>{\rangle}

\begin{document}

\title{Extreme commutative quantum observables are sharp} 

\author{Teiko Heinosaari}
\email{teiko.heinosaari@utu.fi}
\author{Juha-Pekka Pellonp\"a\"a}
\email{juhpello@utu.fi}

\begin{abstract}
It is well known that, in the description of quantum observables, positive operator valued measures (POVMs) generalize projection valued measures (PVMs) and they also turn out be more optimal in many tasks.
We show that a commutative POVM is an extreme point in the convex set of all POVMs if and only if it is a PVM. 
This results implies that non-commutativity is a necessary ingredient to overcome the limitations of PVMs.
\end{abstract}

\maketitle


\section{Introduction}

Quantum observables are generally described by \emph{positive operator valued measures (POVMs)} \cite{QTOS76}. 
In the more traditional description, observables are identified with \emph{projection valued measures (PVMs)}. 
POVMs generalize PVMs in a straightforward way; their elements need not be projections but only positive operators.
PVMs are called \emph{sharp observables} when there is a need to emphasize this distinction \cite{OQP97}.

Even if PVMs turned out to be a too restrictive idealization to describe actual experimental settings, they are still consider to have an exceptional role.
It seems fair to say that not all POVMs represent physical quantities, although they all correspond to some measurement procedures. 
Also, it is quite clear that not all POVMs are useful. 
For instance, coin tossing can be written as a POVM, but this kind of procedure clearly does not give any information on the input state.

For these reasons, there is a motivation to understand further the structure of the set of POVMs.
In particular, it seems useful to identify those intrinsic properties that any useful or optimal POVM must have.
The most evident structure of the set of POVMs is the convex structure \cite{PSAQT82}.
A convex combination of two POVMs describes a mixture of the corresponding measurement apparatuses, similarly as convex combination of two states describes a mixture of the corresponding preparators.

Analogously to pure states, some observables are \emph{extreme} in the sense that they cannot be implemented by mixing some other observables.
It is easy to see that every PVM is an extreme element in the convex set of all POVMs.
However, it is also known that there are other extreme elements than PVMs and this is one the main reasons that motivates one to study POVMs.
Based on earlier works \cite{DaLoPe05}, \cite{Parthasarathy99}, a general characterization of all extreme POVMs was recently derived in \cite{Pellonpaa11}.
However, this (quite technical) result does not reveal qualitative properties of extreme POVMs.

Our contribution in this work is to show that \emph{every extreme commutative POVM is a PVM}.
Therefore, only non-commutative POVMs can outperform PVMs.
This observation explains why all previously found optimal POVMs are always either PVMs or non-commutative POVMs.

In Section \ref{sec:ext} we shortly recall the basics of the convex structure of POVMs.
The main result is stated and proved in Section \ref{sec:commu}.
We give two different proofs of this result.
The first one is a direct proof and applies to discrete POVMs.
The second one starts from the structure of commutative POVMs and applies also to non-discrete POVMs.

\section{Convex set of POVMs}\label{sec:ext}

Let us briefly recall the mathematical description of quantum observables via POVMs \cite{PSAQT82,OQP97}. Consider a quantum system associated with a separable (either finite or infinite dimensional) Hilbert space $\hi$.
The possible measurement outcomes form a set $\Omega$, and $\Sigma$ is any $\sigma$-algebra of subsets of $\Omega$. 
In typical examples, $\Omega$ is the real line $\R$ or its subset, and $\Sigma$ is the Borel $\sigma$-algebra  $\mathcal{B}(\Omega)$.
However, we do not assume any extra structure on $\Omega$ and $\Sigma$.

Let $\lh$ be the set of bounded operators on $\hi$.
A quantum state $\varrho$ is a positive trace class operator of trace one. 
A POVM is set a function $\Ao:\,\Sigma\to\lh$ such that, for every state $\varrho$ the mapping
$X\mapsto \tr{\varrho\Ao(X)}$
is a probability measure. Especially, $\Ao$ satisfies the normalization condition $\Ao(\Omega)=I$ and each operator $\Ao(X)$ satisfies $0 \leq\Ao(X)\leq\id$; operators satisfying this inequality are called \emph{effects}. 
The number $\tr{\varrho\Ao(X)}$ is the probability of getting a measurement outcome $x$ belonging to $X$, when the system is in the state $\varrho$ and the measurement of $\Ao$ is performed. 

We denote by $\Povms$ the set of all POVMs with a fixed measurement outcome space $\Omega$, $\sigma$-algebra $\Sigma\subseteq 2^\Omega$, and Hilbert space $\hi$.
For any pair of POVMs $\Ao,\Bo\in\Povms$, we can define their convex combination POVM $t\Ao+(1-t)\Bo$ (with weight $0<t<1$) by the formula
\begin{equation*}
\bigl( t\Ao+(1-t)\Bo \bigr)(X) =  t\Ao(X)+(1-t)\Bo(X) \, .
\end{equation*}
The convex combination  $t\Ao+(1-t)\Bo$ corresponds to a classical randomization or mixing between $\Ao$ and $\Bo$. 

 A POVM $\Ao\in\Povms$ is called \emph{extreme} if, for all $\Bo,\Co\in\Povms$ and $0<t<1$, the condition  $\Ao= t\Bo+(1-t)\Co$ implies $\Bo=\Co=\Ao$.
Thus, extreme POVMs are free from any classical noise arising from this type of randomization.

A POVM $\Ao\in\Povms$ is a projection valued measure (PVM) if all $\Ao(X)$ are projections, i.e.,
\begin{equation}\label{eq:pvm1}
\Ao(X)^2=\Ao(X) \quad \textrm{for all $X\in\Sigma$} \, .
\end{equation}
All PVMs are extreme since the projections are the extreme elements of the convex set of all effects \cite{QTOS76}. 

There are also extreme POVMs that are not PVMs.
A physically interesting extreme POVM is the canonical phase observable \cite{HePe09}.
It is not a PVM, but is consider to be related to the optimal measurement of the phase of an optical field.

\section{Extreme commutative POVMs}\label{sec:commu}

Suppose $\Ao\in\Povms$ is a PVM. From \eqref{eq:pvm1} we obtain
\begin{equation}\label{eq:pvm2}
\Ao(X)\Ao(Y)=\Ao(X\cap Y) \quad \textrm{for all $X,\,Y\in\Sigma$} \, .
\end{equation} 
It follows that any PVM satisfies
\begin{equation}\label{eq:commu}
\Ao(X)\Ao(Y)=\Ao(Y)\Ao(X) \quad \textrm{for all $X,\,Y\in\Sigma$} \, .
\end{equation}
Generally, a POVM $\Ao$ satisfying \eqref{eq:commu} is called {\it commutative}.

In summary, every PVM is commutative and extreme.
Actually, these two properties characterize PVMs completely.
Our main result is the following.
 
\begin{theorem}
A commutative POVM is extreme if and only if it is PVM.
\end{theorem}

We present two proofs that use totally different techniques.
The first proof is a direct proof and requires no previous results, while the second proof starts from the structure of commutative POVMs.

\subsection*{First proof}
In the first proof restrict to POVMs that are discrete.
Let $\Omega=\{x_1,x_2,\ldots\}$ be a finite or a countably infinite set with $N$ elements (hence $N\in\N$ or $N=\infty$), $\Sigma=2^\Omega=\{X\subseteq\Omega\}$, and $\Ao:\,\Sigma\to\lh$ a POVM. 
We denote $\Ao_n:=\Ao(\{x_n\})$, and the normalization of $\Ao$ then reads $\sum_{n=1}^{N}\Ao_n=I$. Without restricting generality, we can assume that $\Ao_n\ne 0$ for all $n=1,\ldots,N$.

Let $(D_n)_{n=1}^N$ be a sequence of bounded selfadjoint operators such that $\|D_n\|\le 1$ and 
$$
\sum_{n=1}^{N}\sqrt{\Ao_n}D_n\sqrt{\Ao_n}=0 \, .
$$
If $\Ao$ is extreme then $\sqrt{\Ao_n}D_n\sqrt{\Ao_n}=0$ for all $n=1,\ldots,N$. 
Indeed, assume that,
for some $k$, $\sqrt{\Ao_k}D_k\sqrt{\Ao_k}\ne 0$ and define POVMs $\Ao^\pm$ by
\begin{equation*}
\Ao^\pm_n:=\sqrt{\Ao_n}(I\pm D_n)\sqrt{\Ao_n}=\Ao_n\pm\sqrt{\Ao_n}D_n\sqrt{\Ao_n} \, .
\end{equation*}
Then $\Ao=\frac12(\Ao^++\Ao^-)$ and $\Ao_k\ne\Ao_k^\pm$, hence yielding a contradiction.

Let $k\ne l$. 
If $\Ao$ is commutative then $\Ao_k\Ao_l=\Ao_l\Ao_k$, and it follows that
\begin{equation*}
\Ao_k(\Ao_l)^2\Ao_k-\Ao_l(\Ao_k)^2\Ao_l=0 \, .
\end{equation*}
We can write this in an equivalently form
$$
\sum_{n=1}^{N}\sqrt{\Ao_n}D_n\sqrt{\Ao_n}=0
$$
where $D_k:=\sqrt{\Ao_k}(\Ao_l)^2\sqrt{\Ao_k}$, $D_l:=-\sqrt{\Ao_l}(\Ao_k)^2\sqrt{\Ao_l}$, and
$D_n:=0$ if $n\ne k$ and $n\ne l$. Note that, for all $n$, $D_n=D_n^*$ and $\|D_n\|\le 1$.

Hence, if $\Ao$ is extreme and commutative then, for all $k\ne l$, 
$$
\Ao_k(\Ao_l)^2\Ao_k=(\Ao_k\Ao_l)^2=0
$$ 
implying $\Ao_k\Ao_l=0$.
It follows that $\Ao_n=\Ao_n\id=\Ao_n(\sum_m \Ao_m)=\Ao_n^2$ and $\Ao$ is therefore a PVM.

\subsection*{Second proof}
This second proof works for arbitrary POVMs, but requires some background results.
The main ingredient is a representation theorem for commutative POVMs in terms of PVMs.
This type of result is known in many forms, but for our purposes the form presented in \cite{JePuVi08} is most useful.

Let $\Ao\in\Povms$ be commutative. 
By Theorem 4.4 of \cite{JePuVi08} there exist a PVM $\Eo:\,\mathcal B(\R)\to\lh$ and a weak Markov kernel $\nu:\,\R\times\Sigma\to\R$ with respect to $\Eo$
such that 
\begin{equation}\label{eq:coarse}
\Ao(X)=\int_\R\nu(y,X)\Eo(dy)
\end{equation}
for all $X\in\Sigma$. 
The fact that $\nu$ is a weak Marko kernel w.r.t. $\Eo$ means that $y\mapsto\nu(y,X)$ is measurable for each $X\in\Sigma$, $\nu(y,\emptyset)=0$ and $\nu(y,\R)=1$ for $\Eo$-almost all $y\in\R$, and for each sequence $X_n\in\Sigma$ of disjoint sets, we have $\nu(y,\cup_n X_n)=\sum_n \nu(y,X_n)$ for $\Eo$-almost all $y\in\R$.
Physically, \eqref{eq:coarse} means that $\Ao$ is a smearing or a coarse-graining of $\Eo$.

Let us then add the assumption that $\Ao$ is extreme.
As proved in \cite[Theorem 3.3]{JePu07}, an extreme POVM $\Ao$ can be written in the form \eqref{eq:coarse} only if $\nu(y,X)\in\{0,1\}$ for $\Eo$-almost all $y\in\R$. 
Fix $X\in\Sigma$ and define $Y_X:=\{y\in\R\,|\,\nu(y,X)=1\}$. Then $\Ao(X)=\Eo(Y_X)$ is a projection.
Therefore, $\Ao$ is sharp.

\section{Conclusions}\label{sec:conc}

We have proved that \emph{every extreme commutative POVM is a PVM}.
This observation indicates that optimal POVMs fall into two quite different classes; they are either PVMs or non-commutative POVMs.

Our investigation is a step towards understanding the more general question \emph{which POVMs are useful}.

\section*{Acknowledgements}
 
This work has been supported by the Academy of Finland (grant no. 1381359) and The Emil Aaltonen Foundation.

\end{document}